\documentclass[
aps,
prx,
reprint,
superscriptaddress,
amsmath,
amssymb,
longbibliography
]{revtex4-2}

\usepackage[noend]{algpseudocode} % pseudocode style for algorithmicx
\usepackage{lineno}
\usepackage[utf8]{inputenc}
\usepackage{graphicx}
\usepackage{amsfonts}
\usepackage{csquotes}
\usepackage{multirow} 
\usepackage{amsmath}
\usepackage[bb=pazo]{mathalpha}
\usepackage{mathrsfs} 
\usepackage{braket}
\usepackage{xcolor}
\usepackage{bbm}
\usepackage{comment}
\usepackage{enumitem} % for customizing lists
\usepackage{siunitx}
\usepackage[normalem]{ulem}

\usepackage{wrapfig}
\usepackage{hyperref}

\begin{document}

\title{Multi-Parameter Two-Photon Polarimetry at the Quantum Limit}

\author{Joseph~Niblo}
\affiliation{Institute of Photonics and Quantum Sciences, School of Engineering and Physical Sciences, Heriot-Watt University, Edinburgh EH14 4AS, United Kingdom}
\author{Luca~Maggio}
\affiliation{School of Mathematics and Physics, University of Portsmouth, Portsmouth PO1 3HF, United Kingdom}
\affiliation{Quantum Science and Technology Hub, University of Portsmouth, Portsmouth PO1 3FX, United Kingdom}
\author{Russell M. J. Brooks}
\affiliation{Institute of Photonics and Quantum Sciences, School of Engineering and Physical Sciences, Heriot-Watt University, Edinburgh EH14 4AS, United Kingdom}
\author{Joseph~Ho}
\affiliation{Institute of Photonics and Quantum Sciences, School of Engineering and Physical Sciences, Heriot-Watt University, Edinburgh EH14 4AS, United Kingdom}
\author{Vincenzo Tamma}
\affiliation{School of Mathematics and Physics, University of Portsmouth, Portsmouth PO1 3HF, United Kingdom}
\affiliation{Quantum Science and Technology Hub, University of Portsmouth, Portsmouth PO1 3FX, United Kingdom}
\affiliation{Institute of Cosmology and Gravitation, University of Portsmouth, Portsmouth PO1 3FX, United Kingdom}
\author{Alessandro Fedrizzi}
\email{a.fedrizzi@hw.ac.uk}
\affiliation{Institute of Photonics and Quantum Sciences, School of Engineering and Physical Sciences, Heriot-Watt University, Edinburgh EH14 4AS, United Kingdom}

\begin{abstract}

Photonic quantum metrology has demonstrated advantages in precision and resource efficiency for a wide range of applications, with several schemes approaching the fundamental quantum Cramér-Rao precision bound (QCRB).
However, the intrinsic incompatibility of quantum measurements represents a hurdle in extending these advantages to the simultaneous estimation of multiple parameters.
In this paper, we present an experimental protocol approaching the QCRB simultaneously in two polarisation parameters, across a wide range of the parameter space, with as few as $\sim 200$ photon pairs, offering advantages for polarimetric sensing for dim sources such as in X-ray astronomy or photosensitive samples.

\end{abstract}

\maketitle

\section{Introduction} 

Photonic quantum metrology has proven to be an effective tool in a wide range of sensing applications~\cite{polino_photonic_2020, pirandola_advances_2018}.
By employing resources such as coherence and entanglement, quantum sensing schemes have achieved improvements in precision and resource efficiency over their classical counterparts.
Numerous photonic implementations have demonstrated such an advantage~\cite{higgins_entanglement-free_2007, xiang_entanglement-enhanced_2011, harnchaiwat2020tracking, daryanoosh_experimental_2018}. 

Many metrological tasks can be framed as parameter estimation tasks.
In this picture, a probe state $\rho$ undergoes a transformation governed by the parameter of interest $\theta$, which is to be estimated.
After $N$ copies of the evolved state have been measured, the outcomes are input into an estimator function $\tilde\theta$, which returns an estimate of the underlying parameter.
In such a task, the sensitivity of the transformed probe state $\rho_\theta$ to the parameter $\theta$ is given by the quantum Fisher information $H(\rho_\theta)$, which provides a fundamental bound for the achievable precision through the \emph{quantum} Cramér-Rao bound (QCRB).
Quantifying precision by the variance of the estimator $\text{Var}(\tilde\theta)$, this takes the form $\text{Var}(\tilde\theta) \ge \frac{1}{N}H^{-1}$, when unentangled probe states are measured independently~\cite{helstrom_quantum_1969}.

Not all measurements are capable of achieving this precision.
For a given measurement $\mathcal{E}$, the maximum achievable precision is given by the Cramér-Rao bound (CRB) $\text{Var}(\tilde\theta) \ge \frac{1}{N}F^{-1}$, where $F(\rho_\theta, \mathcal{E})$ is the classical Fisher information.
Since $F \le H$, the Fisher information and the quantum Fisher information must coincide for an optimal measurement.
When a single parameter is estimated, it is guaranteed that such a measurement exists~\cite{braunstein_statistical_1994}.
Several photonic schemes have demonstrated operation approaching this precision for example, in estimating the transmission of macroscopic states of light~\cite{woodworth_transmission_2022}, the arrival time of photons on a beam splitter~\cite{PhysRevApplied.19.044068,brooks_optical_2026}, the distance between interfering photons~\cite{PhysRevA.111.032605}, photonic sources~\cite{muratore_superresolution_2025}, and the position of single-photon emitters~\cite{PhysRevLett.132.180802,guo_quantum_2025}.
In many cases, accounting for all resources, including loss or post-selection, is challenging; although some implementations have been shown to approach the QCRB, even when these are taken into account~\cite{allen_approaching_2020, slussarenko_unconditional_2017, zhao_field_2021}.

Recently, a growing interest in more complex sensing scenarios has prompted efforts to extend quantum sensing schemes to the simultaneous estimation of multiple parameters~\cite{pezze_advances_2025}, for example, to measure multiple components of a vector field~\cite{isogawa_entanglement-assisted_2026, li_multiparameter_2026}, or in hybrid sensing tasks like simultaneous magnetometry and thermometry, for applications such as battery monitoring technology~\cite{hatano_simultaneous_2021, ullah_configuration_dependent_2026}.
In photonics, multi-parameter estimation has been demonstrated experimentally on the centroid and separation of incoherent sources of light~\cite{parniak_beating_2018}, phase parameters~\cite{roccia_entangling_2017, cimini_quantum_2019, roccia_multiparameter_2018, cimini_adaptive_2019}, and multiple parameters encoding a unitary evolution of polarisation~\cite{zhou_quantum-enhanced_2015}.

However, the introduction of multiple parameters presents a fundamental hurdle in quantum metrology.
In general, the measurements required to saturate the QCRB for each parameter will be incompatible, and uncertainty relations will forbid simultaneously achieving the maximum precision~\cite{szczykulska_multi-parameter_2016, demkowicz-dobrzanski_multi-parameter_2020}.
Parametrisations allowing simultaneous saturation are therefore of significant interest, and several schemes have been proposed~\cite{maggio_multi-parameter_2025, maggio_ultimate_2026, maggio_quantum-limited_2026}.
In photonics, multi-parameter schemes approaching the QCRB have been demonstrated experimentally for up to three independent phases~\cite{valeri_experimental_2023, cimini_variational_2024}.
However, these demonstrations are limited to a small range of points in parameter space where saturation can be achieved, necessitating adaptive strategies to extend to the full parameter space~\cite{hou_zerotrade-off_2021}.

In a recent work~\cite{maggio_multi-parameter_2025}, the authors demonstrated that multiple polarisation parameters of a two-photon state could, in principle, be estimated simultaneously at the QCRB across a wide range of the parameter 
space.
Quantum polarimetry schemes~\cite{sgobba_single_2023,yoon_experimental_2020} have already demonstrated advantages in the single parameter case and have many natural applications.
For example, in chirality sensing, polarimetric sensing schemes provide rich information about molecular or nanophotonic chiral structures~\cite{hutt_drug_1996, lough_chirality_2002,amabilino_chirality_2009}. 
In this context, quantum schemes operating at the single-photon level are particularly well suited for the study of fragile or photosensitive systems, as reduced optical exposure can minimize sample damage~\cite{kiepas_optimizing_2020, casacio_quantum-enhanced_2021}. 
Such schemes may also be required when polarisation information must be extracted from dim sources.
For example, in astrophysics, X-ray polarimetry of distant objects allows the sensing of anisotropic geometry, magnetic fields, and relativistic effects linked to black hole spin~\cite{costa_efficient_2001, bellazzini_photoelectric_2013, zhang_enhanced_2018, weisskopf_overview_2018}.
Multi-parameter estimation therefore offers a pathway to expand these advantages to higher-dimensional polarisation parameter spaces.

In this work, we demonstrate the simultaneous estimation of two polarisation parameters of a two-photon state.
The scheme proposed in Ref.~\cite{maggio_multi-parameter_2025} does not account for imperfect polarisation states or interference visibility, which are required for a practical implementation.
We therefore extend the framework to include these, making the scheme robust to experimental noise.
We perform the scheme in the many- and few-sample regimes and report good agreement with the updated theory.
Lastly, we demonstrate performance approaching saturation of the QCRB in both parameters simultaneously across a large range of the parameters, with as few as $\sim200$ samples.

\section{Methods}

\begin{figure}
    \includegraphics[width=\linewidth]{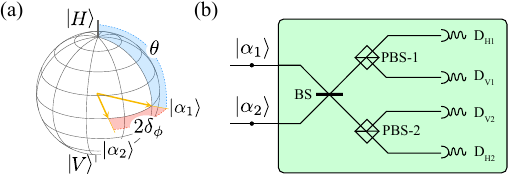}
    \caption{(a) The two parameters being estimated can be represented on a Bloch-Poincar\'{e} sphere for two qubits $\{|\alpha_1\rangle,|\alpha_2\rangle\}$ as state vectors. Here $\theta$ refers to the polar angle of the azimuthal plane shared by the two states (in blue) and $2\delta_\phi$ gives the azimuthal angle between the two states (in red). (b) Illustration of the multi-parameter estimation scheme. The photonic qubits $\{|\alpha_1\rangle,|\alpha_2\rangle\}$ are measured on a polarisation-resolved two-photon interferometer. This consists of a balanced beam splitter (BS) with each output port subject to a polarising beam splitter (PBS) then measured on detectors $\{D_{H1},D_{V1},D_{H2},D_{V2}\}$.
    }
    \label{fig:conceptual}
\end{figure}

Our quantum metrology scheme is defined with respect to two parameters, $\theta$ and $\delta_\phi$, encoded on the two-photon polarisation state $\ket{\psi}=\ket{\alpha_1}\ket{\alpha_2}$, with the polarisation-encoded single-photon states
\begin{align}
    \ket{\alpha_i} =
    \cos\frac{\theta}{2}\ket{H} +
    \sin\frac{\theta}{2} e^{i\phi_i}\ket{V},
    \quad i \in \{1,2\}, \label{eq:nominal-states}
\end{align}
where $\ket{H}$ corresponds to horizontal polarisation and $\ket{V}$ to vertical polarisation.
The states $\ket{\alpha_i}$ have the same polar angle $\theta$, and $\delta_\phi := \left| {\phi_1-\phi_2}\right|/2$ is defined as half the difference in the azimuthal angles $\phi_i$, as depicted on the Bloch sphere in \autoref{fig:conceptual}~(a).

From this parametrisation, the multi-parameter quantum Cramér-Rao bound can be obtained for the variances of any unbiased estimators of $\theta$ and $\delta_\phi$.
With $C = \text{Cov}(\tilde\theta, \tilde\delta_\phi)$ the covariance matrix for estimators $\tilde\theta$ and $\tilde\delta_\phi$, the multi-parameter QCRB is $C \ge \frac{1}{N}H^{-1}$, after $N$ independent measurements~\cite{holevo_probabilistic_2011, paris_quantum_2004}, where $H$ is the quantum Fisher information matrix, which is obtained directly from the parametrised input state as~\cite{maggio_multi-parameter_2025}
\begin{equation}
H = 2 \left( \begin{array}{cc}1 & 0\\
    0 & 
    {\sin }^{2}\theta \end{array} \right).
\end{equation}
Similarly, the multi-parameter CRB is given by $C \ge \frac{1}{N}F^{-1}$, where $F(\ket\psi, \mathcal{E})$ is the Fisher information matrix for a measurement $\mathcal{E}$, and $F=H$ is necessary to saturate the QCRB.

As shown in Ref.~\cite{maggio_multi-parameter_2025}, this can be achieved using two-photon interference, with the optimal measurement implemented as shown in \autoref{fig:conceptual}~(b).
At the central balanced beam splitter (BS), the two photons exit from the same port depending on their mode overlap, given by $\delta_\phi$~\cite{hong_measurement_1987}.
Sensitivity to $\theta$ is achieved by projecting the photons onto $\{\bra{H}, \bra{V}\}$ using polarising beam splitters (PBSs).
From the outcomes of this measurement, the parameters can be estimated using maximum likelihood estimation, which will asymptotically saturate the QCRB as the number of samples increases~\cite{geyer_asymptotics_2013}.

This result is based on the assumption that the two photons are perfectly indistinguishable in all degrees of freedom except for polarisation.
In practice, imperfect spatial overlap at the BS, together with distinguishability in extraneous degrees of freedom such as frequency, will reduce the interference visibility, and the precision of the estimators will be reduced.
To model these imperfections, we introduce an indistinguishability parameter, $\eta \in [0,1]$, for the two input photons at the BS.

The maximum saturation of the QCRB for $\eta\neq1$ is derived in Appendix~\ref{sec:calculating-FIM} from the modified Fisher information matrix as
\begin{equation}
    F'(\theta, \delta_\phi, \eta) = 2 \left( \begin{array}{cc}1 & 0\\
    0 & 
    {\sin }^{2}\theta \frac{\eta^2\sin^22\delta_{\phi}}{1 - \eta^2\cos^22\delta_{\phi}}
    \end{array}\right)
.\end{equation}
Considering the diagonal elements $F'_{\theta\theta}, F'_{\delta_\phi\delta_\phi}$, we see that the CRB for $\text{Var}(\theta)$ is unchanged, and we still have $F'_{\theta\theta} = H_{\theta\theta}$ allowing saturation of the QCRB.
However, for $\delta_\phi$, we have $F'_{\delta_\phi\delta_\phi} \ne H_{\delta_\phi\delta_\phi}$, and the maximum possible saturation of the QCRB is dependent on both $\eta$ and $\delta_\phi$.
The reduction of the Fisher information is least for $\delta_\phi = \pi/4$, where the maximum saturation is bounded by $F'_{\delta_\phi \delta_\phi}/H_{\delta_\phi\delta_\phi} \le \eta^2$.

The $\eta$-dependent maximum likelihood estimators are also obtained in Appendix~\ref{sec:outcome-probabilities}. 
After $N$ independent measurements, these can be written as
\begin{align}
    \tilde{\theta} &= \arccos\left(\frac{N_{HH} - N_{VV}}{N}\right),\label{eq:theta-estimator-nu} \\
    \tilde{\delta}_{\phi} &= \frac{1}{2} \arccos\left( \frac{1}{\eta} \frac{N_{SB} - N_{C'}}{N_{SB} + N_{C'}} \right), \label{eq:delta-phi-estimator-nu}
\end{align} 
where $N_{HH}, N_{VV}, N_{C'}, N_{SB}$ represent the number of occurrences of the following outcomes.
We label events $HH$ when photons are detected only in $H$ detectors, and $VV$ similarly.
In the remaining events, the photons are detected with different polarisations, which we label as $C'$ when the photons are detected in different output ports of the BS, and $SB$ when they are detected in the same.

\begin{figure}
    \centering
    \includegraphics[width=1\linewidth]{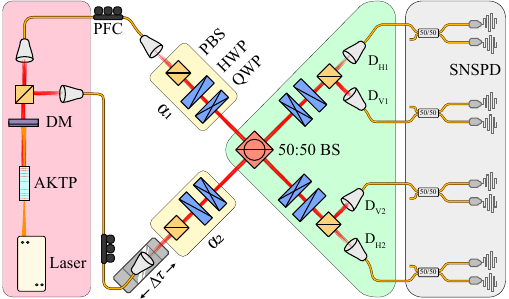}
    \caption{\textbf{Experimental setup.} The red block highlights the two-photon source.
    A laser optically pumps an aperiodically-poled KTP (AKTP) crystal for type-II parametric down-conversion.
    A dichroic mirror (DM) removes the pump laser, while the photon pairs are separated on a polarising beam splitter (PBS) and coupled into single mode fibre.
    Polarisation fibre controllers (PFC) are used to undo unwanted rotations.
    Each yellow block shows the polarisation qubits \{$|\alpha_1\rangle,|\alpha_2\rangle$\} encoded by a PBS, half-wave plate (HWP), and quarter-wave plate (QWP).
    The green triangular block is the polarisation-resolved two-photon interferometer,  consisting of a 50:50 beam splitter cube (50:50 BS) followed by a PBS in each output arm. The additional HWPs and QWPs are used to perform quantum state tomography. Photons are detected using superconducting nanowire single-photon detectors (SNSPD), shown in the grey block. Each output uses an in-fibre 50:50 BS for pseudo-photon number resolving detection, described in the main text.}
    \label{fig:experimental-setup}
\end{figure}

The experimental setup we used to implement the scheme is shown in \autoref{fig:experimental-setup}. 
An aperiodically poled KTP crystal, quasi-phase-matched for Type-II parametric down-conversion, was used to produce photon pairs at 1550~nm~\cite{pickston_optimised_2021, graffitti_independent_2018}.
The crystal was pumped by a pulsed titanium sapphire laser with a repetition rate of 80~MHz and centre wavelength of 775~nm.
Using 40~mW of average optical power, we recorded $\sim10^5$ pairs per second through the setup.
The signal and idler were separated with a PBS and we used a combination of a PBS, followed by a half-wave plate and a quarter-wave plate, to prepare the polarisation states for each photon, defined in \autoref{eq:nominal-states}~\cite{langford_encoding_2007}.
The indistinguishability $\eta = 97.7\%$ was estimated from the visibility obtained by sweeping the delay $\Delta\tau$ between the two photons at the central BS, with both photons prepared in $\ket{H}$.

The probabilistic generation of the two-photon state was heralded by post-selecting on two-photon detection events to exclude single-photon background and events where one of the two photons is lost.
However, some two-photon events result in only a single detector click.
To distinguish these, we implemented pseudo-photon number resolving detection by demultiplexing each output arm to two detectors through an in-fibre 50:50 BS.
This allows, on average, half of the single-detector outcomes to be resolved, so we also removed half of the two-detector outcomes prior to evaluating the estimators to recover the statistics that would be observed with true number-resolving detection.

\section{Results}

\begin{figure} 
   \centering
   \includegraphics[width=\linewidth, 
   trim=8 8 2 2  % left bottom right top (in points)
   ]{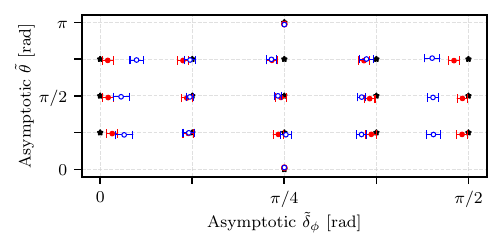}
   \caption{
    \textbf{Many-sample bias.} Nominal state preparations are shown as black stars, QST values are shown as red filled circles and estimator values as unfilled blue circles. Errors in the estimators are estimated from the standard deviation of $\sim 30$ repetitions. Errors for tomography are described in Appendix~\ref{sec:state-tomography}. We omit errors for $\theta$ as they are too small to be clearly seen. At the poles of the Bloch sphere, with $\theta \in \{0, \pi\}$, $\delta_\phi$ is not geometrically meaningful, so we set $\delta_\phi = \pi/4$ for clarity.
    }
   \label{fig:asymptotic-results}
\end{figure}

\begin{figure*} 
   \centering
   \includegraphics[width=\linewidth, 
   trim=14 14 2 2  % left bottom right top (in points)
   ]{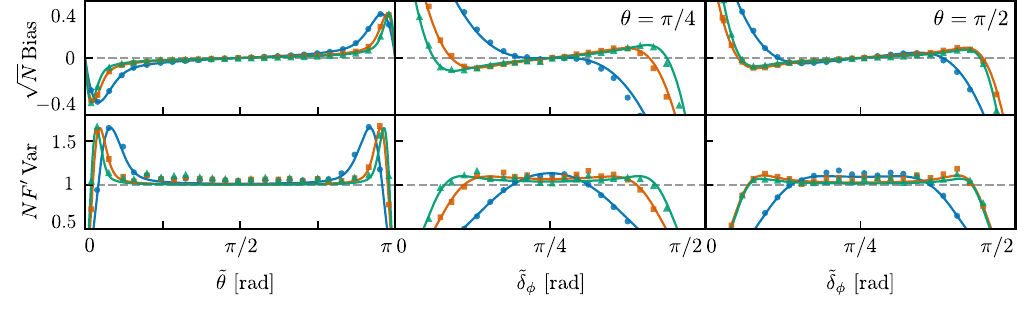}
   \caption{
   \textbf{Few-sample bias and precision.}
   In the top row, the bias of the estimators is shown normalised by $\sqrt N$. 
   In the bottom row, the variance of the estimators is shown, normalised by the Fisher information $NF'$, so that the CRB is normalised to one.
   Two plots are shown for $\tilde \delta_\phi$, for two values of $\theta \in \{\pi/4, \pi/2\}$.
   Experimental values are shown as points and theoretical curves are shown as lines.
   Three values of $N \in \{40, 120, 200\}$ are shown, in blue circles, orange squares and green triangles, respectively. Errors were calculated from the estimator variances but are omitted due to size.
   }
   \label{fig:variance-results}
\end{figure*}

First, we evaluated the protocol in the asymptotic regime.
We prepared 17 nominal states $\{\vert\alpha_1\rangle,\vert\alpha_2\rangle\}$ to span the parameter space in a uniform manner, as shown by the black stars in \autoref{fig:asymptotic-results}.
For each setting, we obtained the estimators using Eqs.~\ref{eq:theta-estimator-nu},~\ref{eq:delta-phi-estimator-nu}, which are shown in the blue empty circles.
The estimators were obtained from $N=2000$ samples, where the bias of the estimators is predicted to be negligible at $< 0.01~\text{rad}$.
We find $\tilde\theta$ shows good agreement with the nominal state preparations across the entire parameter range, up to an average bias of $0.03~\text{rad}$ across all states.
For $\tilde{\delta}_\phi$, the average bias is within $0.03~\text{rad}$ in the central region $\delta_\phi \in [\pi/4, 3\pi/4]$, but greater for extremal values of $\delta_\phi$, with an average bias of $0.13~\text{rad}$ towards less extremal values.

This is partially a result of imperfect state preparations, which we confirmed by performing full quantum state tomography (QST) to reconstruct the density matrices for the polarisation of each photon.
For each state, we recorded $\sim2\times10^5$ events in each of three basis measurements to obtain asymptotic estimates of the ground truth $\theta$ and $\delta_\phi$, shown as red solid circles in \autoref{fig:asymptotic-results}.
The protocol estimators agree more closely with these values, to within $0.02~\text{rad}$ on average for $\theta$ and for $\delta_\phi$ in the central region, and to within experimental errors for $95\%$ of these values.
In the extremal region, the estimator for $\delta_\phi$ still shows an average bias of $0.11~\text{rad}$ towards less extremal values compared to QST.

At the extremal points $\delta_\phi \in \{0,\pi\}$, which are the least and most distinguishable states, respectively, we expect the protocol to be most sensitive to imperfect state preparations.
In particular, when the two states are prepared with a small difference in polar angles $\delta_\theta = |\theta_1 - \theta_2|$, the effect on the estimators can be found analytically, as discussed in Appendix~\ref{sec:delta-theta}.
We observed an average deviation $\bar{\delta_\theta} = 0.04~\text{rad}$ over all prepared states, for which $\tilde\theta$ is not significantly affected, but the effect on $\tilde{\delta}_\phi$ becomes non-negligible for $\delta_\phi \approx 0$ and $\delta_\phi \approx\pi / 2$.
The protocol is robust to small $\delta_\theta$ in the central region for $\delta_\phi$, which coincides with the region where $\tilde\delta_\phi$ is unbiased and most saturates the QCRB, as discussed below.

We now explore the dependence on the number of samples $N$ in the non-asymptotic regime.
For each estimator, we identify the region where it is unbiased, and we compare the variance to the CRB and the QCRB.
We performed the protocol across a range of settings for $(\theta, \delta_\phi)$.
For each setting, we obtained $\sim 70,000$ measurement outcomes, which were grouped for estimation into sets of size $N \in \{40, 120, 200\}$, and the estimators were evaluated.
The bias of each estimator was calculated as the average difference from the asymptotic value to correct for the asymptotic bias discussed above, and the variance was obtained from all experimental repetitions with fixed $N$.
These are shown in \autoref{fig:variance-results}, alongside the theoretical bias and variance discussed in Appendix~\ref{sec:probability-distributions-updated-estimators}.

For $\tilde\theta$, the bias and variance are independent of the value of $\delta_\phi$, so we set $\delta_\phi = 0$ and prepared states across the range $\theta \in [0, \pi]$.
We find that the estimator is unbiased around $\theta = \pi/2$, with the width of the unbiased region increasing with $N$.
The variance of the estimator quickly saturates the CRB with increasing $N$ in the unbiased region, and the width of the saturation region also increases with $N$.
The experimental variance of the estimator agrees with the theory up to a noise floor that is roughly constant across the saturation region, attributable to unaccounted sources of experimental noise.

For $\tilde{\delta}_\phi$, the bias and variance depend on $\theta$, and we report statistics for two values of $\theta \in \{{\pi}/{4}, \pi/2\}$ across the range $\delta_\phi \in  [ 0,\pi/2]$. 
The estimator is least biased for non-extremal values around $\delta_\phi = \pi/4$, but with $\eta < 1$, the gradient of the bias is non-zero in this region.
This gradient is steeper for $\theta = \pi/4$, but in both cases, it is approximately proportional to $1/\sqrt{N}$ and will decay asymptotically.
The variance of the estimator approaches the CRB in the unbiased region for $N\sim 200$.
For fixed $N$, the unbiased region and the saturation region for $\tilde{\delta}_\phi$ are wider for $\theta = {\pi}/{2}$ than for $\theta = {\pi}/{4}$. 
The Fisher information, $F_{\delta_\phi\delta_\phi}$, is also greater for $\theta = {\pi}/{2}$, indicating that both the maximum precision in the estimator for $\delta_\phi$ is greater for $\theta = \pi/2$ and that a greater range of values of $\delta_\phi$ are able to achieve it.

Finally, we calculate the saturation of the QCRB for each parameter, for $N=200$.
Taking the unbiased region for the estimators to be the central half of the parameter range
$
    \tilde{\theta}^\mathrm{ub} \in [\pi/4, 3\pi/4] , \,
    \tilde{\delta}_{\phi} ^{\mathrm{ub}} \in [3\pi/8, 5\pi/8] ,
$
we define the saturation $S := (NH)^{-1} / \left<\text{Var}\right>_{\text{ub}}$ as the minimum possible variance, given by the QCRB, divided by the average variance over the unbiased region.
For $\tilde\theta$, the saturation is limited only by the noise floor in the unbiased region, while for $\tilde\delta_\phi$, the saturation is bounded by $\eta^2 = 95\%$ due to the reduced Fisher information, from imperfect indistinguishability.
We find
\begin{align*}
    S(\tilde\theta)  &= (93 \pm 2 )\%, \\
    S(\tilde\delta_\phi^{\,\theta = \pi/2})  &= (89 \pm 2)\%, \\
    S(\tilde\delta_\phi^{\,\theta = \pi/4})  &= (86 \pm 3)\%.
\end{align*}

\section{Discussion}

In this work, we have demonstrated a photonic multi-parameter estimation scheme approaching the ultimate quantum Cramér-Rao precision bound.
The simultaneous saturation of the CRB was achieved in two polarisation parameters, with as few as $\sim200$ photon pairs.
Saturation was shown across a wide parameter range, which increases quickly with the number of samples.

This work is a natural extension of the estimation scheme proposed in Ref.~\cite{harnchaiwat2020tracking}, where the authors estimate a single polarisation parameter using two-photon interference.
We have shown that the introduction of polarisation-resolving measurements allows for the simultaneous estimation of two polarisation parameters, and that the precision in both parameters can approach the QCRB.
This represents an advantage over standard techniques for resolving the polarisation state of qubits, for example, using quantum state tomography.
As shown in Appendix~\ref{sec:state-tomography}, QST can require up to three times as many photon pairs to achieve the same precision, corresponding to the three \textit{sequential} measurements that are performed to obtain the full two-photon polarisation state.
In contrast, our protocol utilises interference to allow the simultaneous estimation of both parameters, while projecting into only a single polarisation basis.

The degree of saturation of the QCRB was limited by imperfect state preparation, reducing the two-photon interference visibility.
This sensitivity is an unavoidable consequence of incompatibility in multi-parameter estimation.
For example, it is impossible to estimate the polar and azimuthal angles (or indeed any multi-parameter parametrisation) of a single qubit simultaneously at the QCRB~\cite{suzuki_information_2019}.

While we work in the conditional regime, the scheme is not inherently restricted by loss.
Since the protocol is defined on independent input states, the estimators can be evaluated on observed outcomes without modification.
In principle, with the use of high-efficiency, number-resolving detectors and low-loss optics, the scheme could be carried out in the unconditional regime.

In future implementations, adaptive strategies could be employed to extend the scheme~\cite{valeri_experimental_2023}.
For example, by introducing an adaptive unitary rotation to one input arm of the interferometer, the difference in azimuthal angles between the photons could be adjusted to $\delta_\phi \approx \pi / 4$, allowing the scheme to operate near maximum efficiency, even for extremal input states.
A similar strategy may also allow the elimination of small differences in polar angle, $\delta_\theta$, which reduces precision near extremal values of $\delta_\phi$.
It may also be possible to extend the protocol to more parameters, for example, by leveraging higher-dimensional multi-photon interference~\cite{faleo_entanglement-induced_2024}.
Since the advantage derived from simultaneous multi-parameter estimation scales with the number of parameters to be estimated, this could open the way to even greater advantages in precision and resource efficiency. \\

\noindent
\textbf{Acknowledgments}
\newline
This work was supported by the UK Engineering and Physical Sciences Research Council (Grant No. EP/Z533208/1).
L. Maggio acknowledges partial support by Xairos Systems Inc. 
V. Tamma acknowledges partial support from
the Air Force office of Scientific Research under award
number FA8655-23-1-7046.

\noindent
\textbf{Author contributions}
\newline
AF and VT conceived the project. 
JN, RB and JH performed the experiment, JN, RB, and LM analysed the results.
LM and VT developed the theoretical tools used in the analysis.
All authors contributed to writing and revisions of the manuscript.

\vspace{1em}
\noindent
\textbf{Competing interests}
\newline
The authors declare no competing financial or non-financial interests.

\bibliographystyle{apsrev4-2}
\bibliography{bib}

\onecolumngrid 
\clearpage

\appendix
\renewcommand{\sectionautorefname}{Appendix}

\section{Incorporating Imperfect Indistinguishability}\label{sec:outcome-probabilities}
Here we generalise the two-photon state, allowing us to account for imperfect mode overlap on the central beam splitter and imperfect indistinguishability in extraneous degrees of freedom.

\subsection{State Evolution}

In full generality, we define the initial state
\begin{equation}
    \ket{\psi}=\hat{a}^\dagger_1(\vec{\alpha}_1)\hat{d}^\dagger_2(\vec{\alpha}_2)\ket{0}
\end{equation}
where $\vec{\alpha}_j$ defines the Bloch vector that represents the polarization state of the $j$-th photon.
We write $\vec{\alpha}_j$ in terms of polar angles as 
\begin{equation}
    \vec{\alpha}_j=\begin{pmatrix}
        \sin\alpha_j\cos\phi_j\\
        \sin\alpha_j\sin\phi_j\\
        \cos\alpha_j
    \end{pmatrix}.
\end{equation}
Lower indexes of creation operators define the spatial mode of the photon arriving at the beam splitter, from either the first or second input port.

To model partial indistinguishability in extraneous degrees of freedom of the two photons, we have in general that $\hat{d}^\dagger_j$ is not orthogonal to $\hat{a}^\dagger_j$.
By defining $\hat{b}^\dagger_j$ as a suitable orthogonal operator with respect to $\hat{a}^\dagger_j$, that is 
\begin{equation}\label{eq:a-b-commutator}
\bigl[\hat{a}_i(\vec{\alpha}_1), \hat{b}^\dagger_j(\vec{\alpha}_2)\bigr]
= 0,
\end{equation}
we may decompose $\hat{d}^\dagger=\sqrt{\eta}\hat{a}^\dagger_j+\sqrt{1-\eta}\hat{b}^\dagger_j$, with $\eta\in[0,1]$.
With $\eta=1$, we have $\hat{d}^\dagger_j=\hat{a}^\dagger_j$ and the two photons are identical in the extraneous degrees of freedom.
If $\eta=0$, we have $\hat{d}^\dagger_j=\hat{b}^\dagger_j$ and the two photons are fully distinguishable.
The commutation relations between the operators are given as
\begin{align}
\bigl[\hat{a}_i(\vec{\alpha}_1), \hat{a}^\dagger_j(\vec{\alpha}_2)\bigr]
&=
\Bigl[
    \cos\!\left(\frac{\alpha_1}{2}\right)\cos\!\left(\frac{\alpha_2}{2}\right)
    + \sin\!\left(\frac{\alpha_1}{2}\right)\sin\!\left(\frac{\alpha_2}{2}\right)
      \mathrm{e}^{i(\phi_2-\phi_1)}
\Bigr]\delta_{i j},
\\[0.5em]
\bigl[\hat{b}_i(\vec{\alpha}_1), \hat{b}^\dagger_j(\vec{\alpha}_2)\bigr]
&=
\Bigl[
    \cos\!\left(\frac{\alpha_1}{2}\right)\cos\!\left(\frac{\alpha_2}{2}\right)
    + \sin\!\left(\frac{\alpha_1}{2}\right)\sin\!\left(\frac{\alpha_2}{2}\right)
      \mathrm{e}^{i(\phi_2-\phi_1)}
\Bigr]\delta_{i j},
\end{align}
with $i,j = 1, 2$.
It is straightforward to check
\begin{equation}\label{eq:a-a-b-b-commutators}
\bigl|\,[\hat{a}_i(\vec{\alpha}_1), \hat{a}^\dagger_j(\vec{\alpha}_2)]\,\bigr|^2
=
\bigl|\,[\hat{b}_i(\vec{\alpha}_1), \hat{b}^\dagger_j(\vec{\alpha}_2)]\,\bigr|^2
=
\frac{1 + \vec{\alpha}_i \cdot \vec{\alpha}_j}{2}.
\end{equation}

The two input photons impinge on a balanced beam splitter, whose effect on the state $\ket{\psi}$ is described by the operator $\hat{U}_{\mathrm{BS}}$. This operator acts on the injected probe through the map 
$\hat{U}_{\mathrm{BS}}\hat{a}_i\hat{U}^\dagger_{\mathrm{BS}}=\sum_j U_{ij}\hat{a}_j$, 
and $\hat{U}_{\mathrm{BS}}\hat{b}_i\hat{U}^\dagger_{\mathrm{BS}}=\sum_j U_{ij}\hat{b}_j$, where
\begin{equation*}
    U=\frac{1}{\sqrt{2}}\begin{pmatrix}
        1&-1\\1&1
    \end{pmatrix},
\end{equation*}
so that the state at the output of the beam splitter is
\[
    \ket{\psi_{\mathrm{BS}}} = \hat{U}_{\mathrm{BS}} \ket{\psi}
    =\frac{1}{2}\left(
        \hat{a}^\dagger_1(\vec{\alpha}_1)-\hat{a}^\dagger_2(\vec{\alpha}_1)
    \right)
    \left(
        \hat{d}^\dagger_1(\vec{\alpha}_2)+\hat{d}^\dagger_2(\vec{\alpha}_2)
    \right)\ket{0}.
\]
After the beam splitter, the two photons arrive at polarizing beam splitters.
Each polarizing beam splitter projects the photons onto two orthogonal states, defined by $\vec{\beta}_j$ and $-\vec{\beta}_j$, where $j=1,2$ indicates the $j$-th polarizing beam splitter, placed in the $j$-th output port of the beam splitter.
Again, we write $\vec{\beta}_j$ in terms of polar angles as
\begin{equation}
    \vec{\beta}_j=\begin{pmatrix}
        \sin\beta_j\cos\phi_{\beta,j}\\
        \sin\beta_j\sin\phi_{\beta,j}\\
        \cos\beta_j
    \end{pmatrix}.
\end{equation}
In the following section, we evaluate the probabilities of finding the photons in each of the possible detector patterns.

\subsection{Outcome Probabilities}

\paragraph*{Double Bunching Events}
We define double bunching events as those for which both photons are detected in the \textit{same} output port of the \textit{same} polarizing beam splitter. 
The probability of having both photons impinge on the $i$-th polarizing beam splitter and projected onto the state associated with the Bloch vector $\pm\vec{\beta}_i$ is
\begin{align*}
        P_{\pm\vec{\beta}_i,i;\pm\vec{\beta}_i,i}
        &=\left\vert\bra{0}\frac{\hat{a}^2_i(\pm\vec{\beta}_i)}{\sqrt{2}}\ket{\psi_{BS}}\right\vert^2
            +\left\vert\bra{0}\frac{\hat{b}^2_i(\pm\vec{\beta}_i)}{\sqrt{2}}\ket{\psi_{BS}}\right\vert^2
            +\left\vert\bra{0}\hat{a}_i(\pm\vec{\beta}_i)\hat{b}_i(\pm\vec{\beta}_j)\ket{\psi_{BS}}\right\vert^2\\
        &=\frac{\eta}{2}\left\vert\frac{2}{\sqrt{2}}
                [\hat{a}_i(\pm\vec{\beta}_i),\hat{a}^\dagger_i(\vec{\alpha}_1)]
                [\hat{a}_i(\pm\vec{\beta}_i),\hat{a}^\dagger_i(\vec{\alpha}_2)]
            \right\vert^2
            + 0
            + \frac{1-\eta}{2}\left\vert[\hat{a}_i(\pm\vec{\beta}_i),\hat{a}^\dagger_i(\vec{\alpha}_1)][\hat{b}_i(\pm\vec{\beta}_i),\hat{b}^\dagger_i(\vec{\alpha}_2)]\right\vert^2 \\
        &=\frac{1+\eta}{4}\left\vert[
            \hat{a}_i(\pm\vec{\beta}_i),\hat{a}^\dagger_i(\vec{\alpha}_1)][\hat{b}_i(\pm\vec{\beta}_i),\hat{b}^\dagger_i(\vec{\alpha}_2)
        ]\right\vert^2,
\end{align*}
where the sum over terms corresponding to $\langle \hat{a}^2 \rangle, \langle \hat{b}^2 \rangle, \langle \hat{a}\hat{b} \rangle$ represents that we have integrated over the extraneous degrees of freedom.

Using \autoref{eq:a-a-b-b-commutators} we obtain
\begin{equation}
\begin{alignedat}{2}
    &P_{\vec{\beta}_1,1;\,\vec{\beta}_1,1} &&= \frac{1+\eta}{16} \, (1 + \vec{\alpha}_1 \cdot \vec{\beta}_1)\,(1 + \vec{\alpha}_2 \cdot \vec{\beta}_1), \\
    &P_{\vec{\beta}_2,2;\,\vec{\beta}_2,2} &&= \frac{1+\eta}{16} \, (1 + \vec{\alpha}_1 \cdot \vec{\beta}_2)\,(1 + \vec{\alpha}_2 \cdot \vec{\beta}_2), \\
    &P_{-\vec{\beta}_1,1;\,-\vec{\beta}_1,1} &&= \frac{1+\eta}{16} \, (1 - \vec{\alpha}_1 \cdot \vec{\beta}_1)\,(1 - \vec{\alpha}_2 \cdot \vec{\beta}_1), \\
    &P_{-\vec{\beta}_2,2;\,-\vec{\beta}_2,2} &&= \frac{1+\eta}{16} \, (1 - \vec{\alpha}_1 \cdot \vec{\beta}_2)\,(1 - \vec{\alpha}_2 \cdot \vec{\beta}_2).
\end{alignedat}
\end{equation}

\paragraph*{Single Bunching Events}

We define single bunching events as those for which the two photons are detected at the \textit{same} polarizing beam splitter but \textit{different} output ports. That is, if the photons arrive at the $i$-th polarizing beam splitter, one photon is projected onto the state corresponding to $\vec{\beta}_i$, and the other to $-\vec{\beta}_i$. The probability is
\begin{align}
    \begin{split}
        P_{\vec{\beta}_i,i;-\vec{\beta}_i,i}&=\left\vert\bra{0}a_i(\vec{\beta}_i)a_i(-\vec{\beta}_i)\ket{\psi_{BS}}\right\vert^2+\left\vert\bra{0}a_i(\vec{\beta}_i)b_i(-\vec{\beta}_i)\ket{\psi_{BS}}\right\vert^2\\
        & \quad+\left\vert\bra{0}b_i(\vec{\beta}_i)a_i(-\vec{\beta}_i)\ket{\psi_{BS}}\right\vert^2+\left\vert\bra{0}b_i(\vec{\beta}_i)b_i(-\vec{\beta}_i)\ket{\psi_{BS}}\right\vert^2 \\
        &=\frac{\eta}{4}\left\vert[a_i(\vec{\beta}_i),a_i(\vec{\alpha}_1)][a_i(-\vec{\beta}_i),a_i(\vec{\alpha}_2)]+[a_i(-\vec{\beta}_i),a_i(\vec{\alpha}_1)][a_i(\vec{\beta}_i),a_i(\vec{\alpha}_2)]\right\vert^2\\
        & \quad+\frac{1-\eta}{4}\left(\left\vert[a_i(\vec{\beta}_i),a_i(\vec{\alpha}_1)][b_i(-\vec{\beta}_i),b_i(\vec{\alpha}_2)]\right\vert^2+\left\vert[a_i(-\vec{\beta}_i),a_i(\vec{\alpha}_1)][b_i(\vec{\beta}_i),b_i(\vec{\alpha}_2)]\right\vert^2\right)+0 \\
        &=\frac{1}{4}\left(\left\vert[a_i(\vec{\beta}_i),a_i(\vec{\alpha}_1)][a_i(-\vec{\beta}_i),a_i(\vec{\alpha}_2)]\right\vert^2+\left\vert[a_i(-\vec{\beta}_i),a_i(\vec{\alpha}_1)][a_i(\vec{\beta}_i),a_i(\vec{\alpha}_2)]\right\vert^2\right) \\
        &\quad +\frac{\eta}{2}\mathrm{Re}\left([a_i(\vec{\beta}_i),a_i(\vec{\alpha}_1)][a_i(-\vec{\beta}_i),a_i(\vec{\alpha}_2)][a_i(-\vec{\beta}_i),a_i(\vec{\alpha}_1)]^*[a_i(\vec{\beta}_i),a_i(\vec{\alpha}_2)]^*\right).
    \end{split}
\end{align}
Therefore, we have
\begin{align}
    \begin{split}
        P_{\vec{\beta}_1,1;-\vec{\beta}_1,1}&=\frac{1}{8}\left[1-(\vec{\alpha}_1\cdot\vec{\beta}_1)(\vec{\alpha}_2\cdot\vec{\beta}_1)+\eta(\vec{\beta_1}\times\vec{\alpha}_1)\cdot(\vec{\beta_1}\times\vec{\alpha}_2)\right],\\
        P_{\vec{\beta}_2,2;-\vec{\beta}_2,2}&=\frac{1}{8}\left[1-(\vec{\alpha}_1\cdot\vec{\beta}_2)(\vec{\alpha}_2\cdot\vec{\beta}_2)+\eta(\vec{\beta_2}\times\vec{\alpha}_1)\cdot(\vec{\beta_2}\times\vec{\alpha}_2)\right].
    \end{split}
\end{align}

\paragraph*{Coincidence Events}

We define coincidence events as those for which the two photons are detected in different output ports of the central balanced beam splitter.
The probability of a coincidence event is
\begin{align}
\begin{split}
P_{\vec{\beta}_i,i;-\vec{\beta}_i,i} &= 
    \left\vert 
        \bra{0} \hat{a}_i(\vec{\beta}_i) \hat{a}_i(-\vec{\beta}_i) \ket{\psi_{BS}}
    \right\vert^2
    + \left\vert 
        \bra{0} \hat{a}_i(\vec{\beta}_i) \hat{b}_i(-\vec{\beta}_i) \ket{\psi_{BS}}
    \right\vert^2 \\
& \quad
    + \left\vert 
        \bra{0} \hat{b}_i(\vec{\beta}_i) \hat{a}_i(-\vec{\beta}_i) \ket{\psi_{BS}}
    \right\vert^2
    + \left\vert 
        \bra{0} \hat{b}_i(\vec{\beta}_i) \hat{b}_i(-\vec{\beta}_i) \ket{\psi_{BS}}
    \right\vert^2 \\
&= 
    \frac{\eta}{4} 
    \Big\vert 
        [\hat{a}_i(\vec{\beta}_i), \hat{a}_i(\vec{\alpha}_1)]
        [\hat{a}_i(-\vec{\beta}_i), \hat{a}_i(\vec{\alpha}_2)] 
        + [\hat{a}_i(-\vec{\beta}_i), \hat{a}_i(\vec{\alpha}_1)]
        [\hat{a}_i(\vec{\beta}_i), \hat{a}_i(\vec{\alpha}_2)]
    \Big\vert^2 \\
& \quad
    + \frac{1-\eta}{4} 
    \Big(
        \left\vert [\hat{a}_i(\vec{\beta}_i), \hat{a}_i(\vec{\alpha}_1)]
                    [\hat{b}_i(-\vec{\beta}_i), \hat{b}_i(\vec{\alpha}_2)] \right\vert^2
        + \left\vert [\hat{a}_i(-\vec{\beta}_i), \hat{a}_i(\vec{\alpha}_1)]
                        [\hat{b}_i(\vec{\beta}_i), \hat{b}_i(\vec{\alpha}_2)] \right\vert^2
    \Big) \\
&= 
    \frac{1}{4} 
    \Big(
        \left\vert [\hat{a}_i(\vec{\beta}_i), \hat{a}_i(\vec{\alpha}_1)]
                    [\hat{a}_i(-\vec{\beta}_i), \hat{a}_i(\vec{\alpha}_2)] \right\vert^2
        + \left\vert [\hat{a}_i(-\vec{\beta}_i), \hat{a}_i(\vec{\alpha}_1)]
                        [\hat{a}_i(\vec{\beta}_i), \hat{a}_i(\vec{\alpha}_2)] \right\vert^2
    \Big) \\
& \quad
    + \frac{\eta}{2} 
    \mathrm{Re} \Big(
        [\hat{a}_i(\vec{\beta}_i), \hat{a}_i(\vec{\alpha}_1)]
        [\hat{a}_i(-\vec{\beta}_i), \hat{a}_i(\vec{\alpha}_2)]
        [\hat{a}_i(-\vec{\beta}_i), \hat{a}_i(\vec{\alpha}_1)]^*
        [\hat{a}_i(\vec{\beta}_i), \hat{a}_i(\vec{\alpha}_2)]^*
    \Big).
\end{split}
\end{align}

Therefore, the four possible coincidence events have the following probabilities to occur
\begin{align}
\begin{split}
P_{\vec{\beta}_1,1;\vec{\beta}_2,2} &= 
    \frac{1-\eta}{16} 
    \Big[
        (1+\vec{\alpha}_1\cdot\vec{\beta}_1)(1+\vec{\alpha}_2\cdot\vec{\beta}_2)
        + (1+\vec{\alpha}_1\cdot\vec{\beta}_2)(1+\vec{\alpha}_2\cdot\vec{\beta}_1)
    \Big] \\
& \quad
    + \frac{\eta}{16} 
    \Big[
        (1-\vec{\alpha}_1\cdot\vec{\alpha}_2)(1-\vec{\beta}_1\cdot\vec{\beta}_2)
    \Big], \\[2mm]
P_{\vec{\beta}_1,1;-\vec{\beta}_2,2} &= 
    \frac{1-\eta}{16} 
    \Big[
        (1+\vec{\alpha}_1\cdot\vec{\beta}_1)(1-\vec{\alpha}_2\cdot\vec{\beta}_2)
        + (1-\vec{\alpha}_1\cdot\vec{\beta}_2)(1+\vec{\alpha}_2\cdot\vec{\beta}_1)
    \Big] \\
& \quad
    + \frac{\eta}{16} 
    \Big[
        (1-\vec{\alpha}_1\cdot\vec{\alpha}_2)(1+\vec{\beta}_1\cdot\vec{\beta}_2)
    \Big], \\[2mm]
P_{-\vec{\beta}_1,1;\vec{\beta}_2,2} &= 
    \frac{1-\eta}{16} 
    \Big[
        (1-\vec{\alpha}_1\cdot\vec{\beta}_1)(1+\vec{\alpha}_2\cdot\vec{\beta}_2)
        + (1+\vec{\alpha}_1\cdot\vec{\beta}_2)(1-\vec{\alpha}_2\cdot\vec{\beta}_1)
    \Big] \\
& \quad
    + \frac{\eta}{16} 
    \Big[
        (1-\vec{\alpha}_1\cdot\vec{\alpha}_2)(1+\vec{\beta}_1\cdot\vec{\beta}_2)
    \Big], \\[2mm]
P_{-\vec{\beta}_1,1;-\vec{\beta}_2,2} &= 
    \frac{1-\eta}{16} 
    \Big[
        (1-\vec{\alpha}_1\cdot\vec{\beta}_1)(1-\vec{\alpha}_2\cdot\vec{\beta}_2)
        + (1-\vec{\alpha}_1\cdot\vec{\beta}_2)(1-\vec{\alpha}_2\cdot\vec{\beta}_1)
    \Big] \\
& \quad
    + \frac{\eta}{16} 
    \Big[
        (1-\vec{\alpha}_1\cdot\vec{\alpha}_2)(1-\vec{\beta}_1\cdot\vec{\beta}_2)
    \Big].
\end{split}
\end{align}

\subsection{Grouping Events}

So far the state evolution has been carried out in full generality, with no restriction on either the polarisations of the two photons or on the projection in each of the polarising beam splitters.
However, the sensing scheme is defined on input photons having the same polar angle and with the polarising beam splitters in the same horizontal plane.
Therefore, we impose
\begin{gather}
\alpha_1=\alpha_2=\theta, \label{eq:same-polar-angle} \\
\vec{\beta}_1=\vec{\beta}_2=\hat{z}=(0,0,1)^T.
\end{gather}
Defining $\phi_2-\phi_1=2\delta_\phi$, the probabilities take the form
\begin{equation}
    \begin{alignedat}{3}
        &P_{\hat{z},1;\hat{z},1}        &&\equiv P_{H,1;H,1} &&=\frac{1+\eta}{16}(1+\cos\theta)^2,\\
         &P_{\hat{z},2;\hat{z},2}       &&\equiv P_{H,2;H,2} &&=\frac{1+\eta}{16}(1+\cos\theta)^2,\\
          &P_{-\hat{z},1;-\hat{z},1}    &&\equiv P_{V,1;V,1} &&=\frac{1+\eta}{16}(1-\cos\theta)^2,\\
          &P_{-\hat{z},2;-\hat{z},2}    &&\equiv P_{V,2;V,2} &&=\frac{1+\eta}{16}(1-\cos\theta)^2,\\
          &P_{\hat{z},1;-\hat{z},1}     &&\equiv P_{H,1;V,1} &&=\frac{1}{8}\sin^2\theta(1+\eta\cos2\delta_\phi),\\
          &P_{\hat{z},2;-\hat{z},2}     &&\equiv P_{H,2;V,2} &&=\frac{1}{8}\sin^2\theta(1+\eta\cos2\delta_\phi),\\
          &P_{\hat{z},1;\hat{z},2}      &&\equiv P_{H,1;H,2} &&=\frac{1-\eta}{8}(1+\cos\theta)^2,\\
          &P_{-\hat{z},1;-\hat{z},2}    &&\equiv P_{V,1;V,2} &&=\frac{1-\eta}{8}(1-\cos\theta)^2,\\
          &P_{\hat{z},1;-\hat{z},2}     &&\equiv P_{H,1;V,2} &&=\frac{1}{8}\sin^2\theta(1-\eta\cos2\delta_\phi),\\
          &P_{-\hat{z},1;\hat{z},2}     &&\equiv P_{V,1;H,2} &&=\frac{1}{8}\sin^2\theta(1-\eta\cos2\delta_\phi).\\
    \end{alignedat}
\end{equation}
For $\eta=1$, we obtain the results of \cite{maggio_multi-parameter_2025}.
To simplify the estimation of $\theta$ and $\delta_\phi$, we define events $HH, VV, SB, C'$ as in the main text, giving
\begin{equation}\label{eq:outcome-probabilities}
    \begin{alignedat}{3}
        &P_{HH}  &&\equiv P_{H,1;H,1}+P_{H,2;H,2}+P_{H,1;H,2}     &&=\frac{1}{4}(1+\cos\theta)^2,\\
        &P_{VV}  &&\equiv P_{V,1;V,1}+P_{V,2;V,2}+P_{V,1;V,2}     &&=\frac{1}{4}(1-\cos\theta)^2,\\
        &P_{SB}  &&\equiv P_{H,1;V,1}+P_{H,2;V,2}                 &&=\frac{1}{4}\sin^2\theta(1+\eta\cos2\delta_\phi),\\
        &P_{C'}  &&\equiv P_{H,1;V,2}+P_{H,1;V,2}                 &&=\frac{1}{4}\sin^2\theta(1-\eta\cos2\delta_\phi).
    \end{alignedat}
\end{equation}

We note that the probabilities $P_{HH}$ and $P_{VV}$ are independent of $\eta$.
In these events, both photons are projected onto the same polarization, so that their interference is only due to indistinguishability in extraneous degrees of freedom.
Therefore, the functional dependence on polarization can be factorised and dependence on $\eta$ is integrated out.
For events in which the photons are projected onto different polarisations, corresponding to $P_{SB}$ and $P_{C'}$, indistinguishability is no longer attributable only to extraneous degrees of freedom, but also to $\delta_\phi$. 
This means that functional dependence on $\eta$ cannot be factorised, although it is still possible to factorise $\theta$, since the dependence is identical for both photons.
Due to the symmetry of the beam splitter dynamics, $P_{H,1;V,2}=P_{V,1;H,2}$ and $P_{H,1;V,1}=P_{H,2;V,2}$, so that it is possible to treat $P_{H,1;V,2}$ and $P_{V,1;H,2}$ ($P_{H,1;V,1}$ and $P_{H,2;V,2}$) as corresponding to a single event. 

\subsection{Maximum Likelihood Estimators}

We consider the scenario of $N$ events, of which $N_X$ occur in each outcome, $X = HH, VV, SB, C'$.
By definition \cite{fisher_mathematical_1922, cramer_mathematical_1999}, the maximum likelihood estimators, $\tilde{\theta}, \tilde{\delta}_\phi$, are given by
\begin{equation}
\begin{aligned}
        &0=\sum_X N_X \frac{\partial\log P_X}{\partial \theta}\vert_{\theta=\tilde{\theta}}, \\
        &0=\sum_X N_X\frac{\partial\log P_X}{\partial \delta_\phi}\vert_{\delta_\phi=\tilde{\delta}_\phi}.
\end{aligned}
\end{equation}
Calculating the partial derivatives explicitly, we obtain for $\theta$,
\begin{equation}    
\sum_X N_X\frac{\partial\log P_X}{\partial \theta}\vert_{\theta=\tilde{\theta}} 
    =-N_{HH}(1-\cos\tilde{\theta})\sin\tilde{\theta}+N_{VV}(1+\cos\tilde{\theta})\sin\tilde{\theta}+(N-N_{HH}-N_{VV})\sin\tilde{\theta}\cos\tilde{\theta},
\end{equation}
and for $\phi$,
\begin{equation}
    \sum_X N_X\frac{\partial\log P_X}{\partial \delta_\phi}\vert_{\delta_\phi=\tilde{\delta}_\phi} 
      = \frac{N_{SB}}{1+\eta\cos 2\tilde{\delta}_\phi}-\frac{N_{C'}}{1-\eta\cos 2\tilde{\delta}_\phi}, 
\end{equation}
which lead to 
\begin{equation}
\begin{aligned}
    \tilde{\theta} &=\arccos\left(\frac{N_{HH}-N_{VV}}{N}\right), \\
        \tilde{\delta}_\phi &= \frac{1}{2}\arccos\left(\frac{1}{\eta}\frac{N_{SB}-N_{C'}}{N_{SB}+N_{C'}}\right).
\end{aligned}
\end{equation}

\section{Fisher Information Matrix}\label{sec:calculating-FIM}

Employing the definition of the Fisher information matrix (FIM) \cite{holevo_probabilistic_2011, helstrom_quantum_1969}, for the outcomes given above, it is calculated as
\[
F'(\theta, \delta_\phi, \eta) = \sum_X F'_X(\theta, \delta_\phi, \eta),
\]
with $X = {HH}, VV, SB, C'$ representing possible outcomes and where 
\[
F'_X = \frac{1}{P_X} \begin{pmatrix}
    (\partial_\theta P_X)^2 & \partial_\theta P_X \partial_{\delta_\phi} P_X \\
     \partial_\theta P_X \partial_{\delta_\phi} P_X & (\partial_{\delta_\phi} P_X)^2
\end{pmatrix},
\]
is a term corresponding to the outcomes $X$ and $P_X$ are the outcome probabilities given in \autoref{eq:outcome-probabilities}.
This may be explicitly calculated to give 
\begin{equation}
    F'(\theta,\delta_\phi, \eta)
    =\begin{pmatrix}
        F'_{\theta\theta}&F'_{\theta\delta_\phi}\\
        F'_{\delta_\phi\theta}&F'_{\delta_\phi\delta_\phi}
    \end{pmatrix}=2\begin{pmatrix}
        1&0\\0&\sin^2\theta\frac{\eta^2\sin^22\delta_\phi}{1-\eta^2\cos^22\delta_\phi}
    \end{pmatrix}.
\end{equation}
By comparison with the FIM in \cite{maggio_multi-parameter_2025}, we notice that the visibility, $\eta$, does not change the information for $\theta$.
This may be attributed to the fact that the functional dependence of probabilities on $\theta$ is unchanged and factorises for every outcome. 

\begin{figure}
    \centering
    \includegraphics[width=0.5\linewidth]{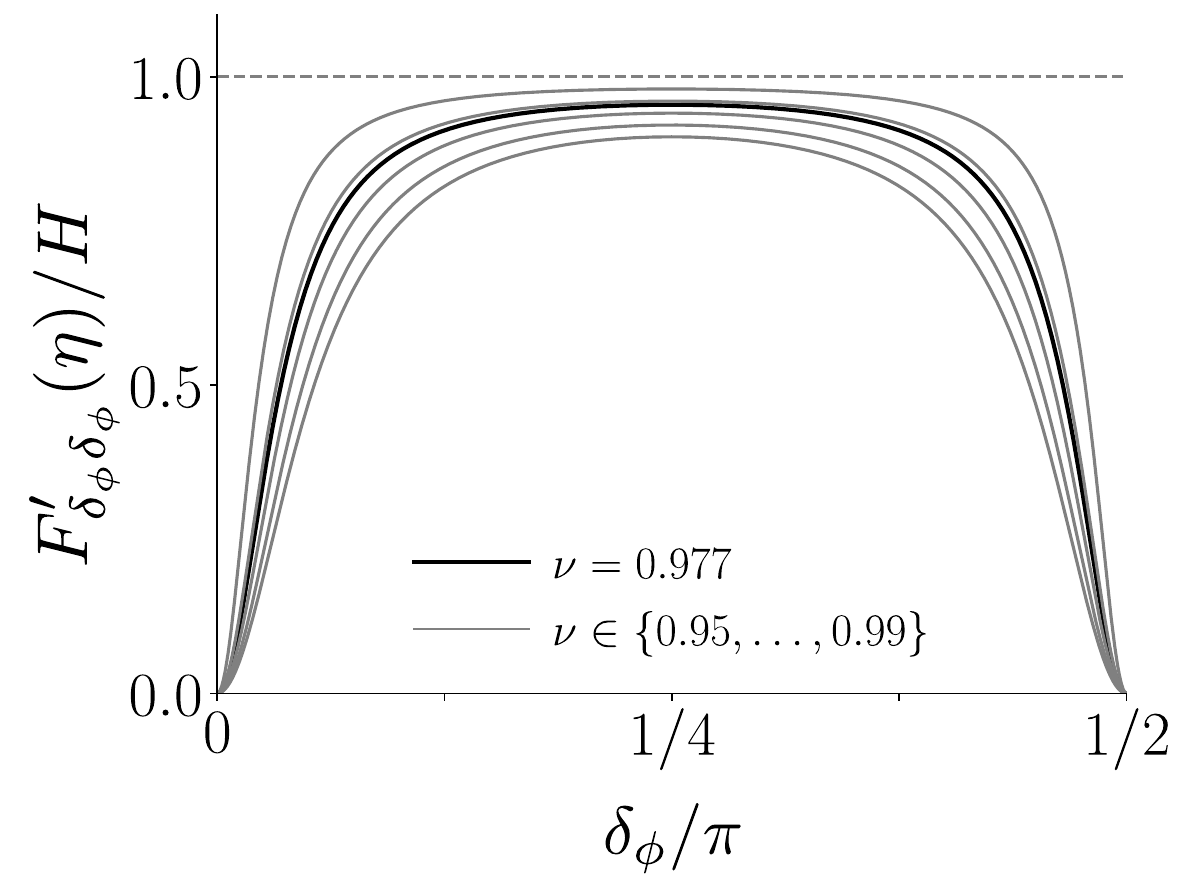}
    \caption{\textbf{Fisher information of $\tilde{\delta}_\phi$ as a function of visibility, $\eta$. } The introduction of an imperfect indistinguishability at the central beam splitter degrades the maximum possible precision of the protocol, especially at extremal values of $\delta_\phi$. The maximum saturation of the QCRB occurs at $\delta_\phi = \pi / 4$.}
    \label{fig:fisher-information-nu}
\end{figure}

From \cite{maggio_multi-parameter_2025}, the quantum Fisher information matrix (QFIM) for the perfect visibility state, with $\eta = 1$, is 
\begin{equation}
    H(\theta, \delta_\phi) = 
    2\begin{pmatrix}
        1 & 0 \\
        0 & \sin^2 \theta
    \end{pmatrix}.
\end{equation}
The Fisher information for $\delta_\phi$, $F'_{\delta_\phi \delta_\phi}$, is compared to the corresponding QFIM element in \autoref{fig:fisher-information-nu}. 
It can be seen that for $\eta < 1$, the Fisher information now depends on $\delta_\phi$ and drops off rapidly at extremal values of $\delta_\phi$.
The maximum occurs at $\delta_\phi = \pi/4$, with a value $\eta^2$.

\section{\texorpdfstring{Introduction of $\delta_\theta$}{Introduction of Difference in Polar Angle}}\label{sec:delta-theta}
In practice, due to inaccuracies in the state preparation, the condition of the two photons to have the same polar angle, \autoref{eq:same-polar-angle}, may not hold exactly.
Here we consider a small deviation, $\delta_\theta$, defined through
\begin{align}
    \alpha_1&=\theta-\delta_\theta/2, \\
    \alpha_2&=\theta+\delta_\theta/2.
\end{align}
The vector $\vec{\alpha}_j$ is then given in terms of $\delta_\theta$ as
\begin{align}
    \vec{\alpha}_j
    &=
    \begin{pmatrix}
        \sin\!\left(
            \theta + \frac{1}{2}(-1)^j \delta_\theta
        \right)\cos\phi_j \\[0.5ex]
        \sin\!\left(
            \theta + \frac{1}{2}(-1)^j \delta_\theta
        \right)\sin\phi_j \\[0.5ex]
        \cos\!\left(
            \theta + \frac{1}{2}(-1)^j \delta_\theta
        \right)
    \end{pmatrix} \\[1ex]
    &=
    \begin{pmatrix}
        \sin\theta \cos\frac{\delta_\theta}{2}\cos\phi_j
        + (-1)^j \cos\theta \sin\frac{\delta_\theta}{2}\cos\phi_j \\[0.5ex]
        \sin\theta \cos\frac{\delta_\theta}{2}\sin\phi_j
        + (-1)^j \cos\theta \sin\frac{\delta_\theta}{2}\sin\phi_j \\[0.5ex]
        \cos\theta \cos\frac{\delta_\theta}{2}
        - (-1)^j \sin\theta \sin\frac{\delta_\theta}{2}
    \end{pmatrix} \\[1ex]
    &=
    \cos\frac{\delta_\theta}{2}
    \begin{pmatrix}
        \sin\theta\cos\phi_j \\[0.5ex]
        \sin\theta\sin\phi_j \\[0.5ex]
        \cos\theta
    \end{pmatrix}
    + (-1)^j \sin\frac{\delta_\theta}{2}
    \begin{pmatrix}
        \cos\theta\cos\phi_j \\[0.5ex]
        \cos\theta\sin\phi_j \\[0.5ex]
        -\sin\theta
    \end{pmatrix} \\[1ex]
    &=
    \cos\frac{\delta_\theta}{2}\,\vec{\alpha}^{\bar\theta}_{j}
    + (-1)^j \sin\frac{\delta_\theta}{2}\,\vec{\alpha}_{j}^\perp,
\end{align}
where in the last line we have defined $\vec{\alpha}_j^{\bar\theta}$ as the unit vector with the same azimuthal angle as $\vec{\alpha}_j$ but the average polar angle over $\vec{\alpha}_1$ and $\vec{\alpha}_2$; and $\vec{\alpha}_{j}^\perp$ as the vector perpendicular to $\vec{\alpha}_j^{\bar\theta}$ in the direction of increasing $\theta$. 

With the measurements still projecting onto $H, V$, the outcome probabilities become
\begin{subequations}
\begin{alignat}{2}
        &P_{\vec{\beta}_1,1;\vec{\beta}_1,1} &&=\frac{1+\eta}{16}\left(\cos\frac{\delta_\theta}{2}+\cos\theta\right)^2, \\
        &P_{\vec{\beta}_2,2;\vec{\beta}_2,2} &&=\frac{1+\eta}{16}\left(\cos\frac{\delta_\theta}{2}+\cos\theta\right)^2, \\
        &P_{-\vec{\beta}_1,1;-\vec{\beta}_1,1} &&=\frac{1+\eta}{16}\left(\cos\frac{\delta_\theta}{2}-\cos\theta\right)^2, \\
        &P_{-\vec{\beta}_2,2;-\vec{\beta}_2,2} &&=\frac{1+\eta}{16}\left(\cos\frac{\delta_\theta}{2}-\cos\theta\right)^2, \\
        &P_{\vec{\beta}_1,1;\vec{\beta}_2,2} &&=\frac{1-\eta}{8}\left(\cos\frac{\delta_\theta}{2}+\cos\theta\right)^2, \\
        &P_{-\vec{\beta}_1,1;-\vec{\beta}_2,2} &&=\frac{1-\eta}{8}\left(\cos\frac{\delta_\theta}{2}-\cos\theta\right)^2, \\
        &P_{\vec{\beta}_1,1;-\vec{\beta}_1,1} &&=\frac{1}{8}\left[2\sin^2\frac{\delta_\theta}{2}+\left(\sin^2\theta-\sin^2\frac{\delta_\theta}{2}\right)(1+\eta\cos2\delta_\phi)\right], \\
        &P_{\vec{\beta}_2,2;-\vec{\beta}_2,2} &&=\frac{1}{8}\left[2\sin^2\frac{\delta_\theta}{2}+\left(\sin^2\theta-\sin^2\frac{\delta_\theta}{2}\right)(1+\eta\cos2\delta_\phi)\right], \\
        &P_{\vec{\beta}_1,1;-\vec{\beta}_2,2} &&=\frac{1}{8}\left[2\sin^2\frac{\delta_\theta}{2}+\left(\sin^2\theta-\sin^2\frac{\delta_\theta}{2}\right)(1-\eta\cos2\delta_\phi)\right], \\
        &P_{-\vec{\beta}_1,1;\vec{\beta}_2,2} &&=\frac{1}{8}\left[2\sin^2\frac{\delta_\theta}{2}+\left(\sin^2\theta-\sin^2\frac{\delta_\theta}{2}\right)(1-\eta\cos2\delta_\phi)\right].
\end{alignat}
\end{subequations}
Therefore, we have
\begin{align}
    \begin{split}
        P_{HH}&=\frac{1}{4}\left(\cos\frac{\delta_\theta}{2}+\cos\theta\right)^2\\
        P_{VV}&=\frac{1}{4}\left(\cos\frac{\delta_\theta}{2}-\cos\theta\right)^2\\
        P_{SB}&=\frac{1}{4}\left[\sin^2\frac{\delta_\theta}{2}(1-\eta\cos2\delta_\phi)+\sin^2\theta(1+\eta\cos2\delta_\phi)\right]\\
        P_{C'}&=\frac{1}{4}\left[\sin^2\frac{\delta_\theta}{2}(1+\eta\cos2\delta_\phi)+\sin^2\theta(1-\eta\cos2\delta_\phi)\right]
    \end{split}
\end{align}
and it follows that the estimator for $\theta$, defined above, now returns
\begin{equation}
    \tilde{\theta}=\arccos\left(\frac{N_{HH}-N_{VV}}{N}\right)\rightarrow\arccos\left(\cos\theta\cos\frac{\delta_\theta}{2}\right).
\end{equation}
For small $\delta_\theta$, $\cos\frac{\delta_\theta}{2} \approx 1$ to first order, so this does not represent a significant change in the estimator.

For the estimator of $\delta_\phi$,
\begin{equation}
    \tilde{\delta}_\phi=\frac{1}{2}\arccos\left(
        \frac{1}{\eta}
        \frac{
            N_{SB}-N_{C'}
        }{
            N_{SB}+N_{C'}
        }
    \right)
        \rightarrow
    \frac{1}{2}
    \arccos\left(
        \frac{
            \sin^2\theta-\sin^2\delta_\theta/2
        }{
            \sin^2\theta+\sin^2\delta_\theta/2
        }
        \cos2\delta_\phi
    \right),
\end{equation}
so that the introduction of $\delta_\theta$ is more significant. 
In particular, the difference is most significant at extremal values of $\delta_{\phi}$, since the factor 
\[ 
\frac{
            \sin^2\theta-\sin^2\delta_\theta/2
        }{
            \sin^2\theta+\sin^2\delta_\theta/2
        }
\]
contributes the greatest change to the argument of the $\arccos$ when $|\cos {2\delta_{\phi}}|$ is greatest.

\section{Probability distributions for the updated estimators}\label{sec:probability-distributions-updated-estimators}

In order to plot the curves in~\autoref{fig:variance-results}, it is necessary to find the probability distributions for the estimators.
In general, given a certain number $N$ of sampling measurements, the outcomes will follow a multinomial distribution, with the probabilities found in the previous section.
The bias and variance of $\tilde\theta$ and $\tilde\delta_{\phi}$ can then be found in the same way as described in Appendix E of Ref.~\cite{maggio_multi-parameter_2025}.
However, it is first necessary to normalise the probability distribution given the possibility that estimation fails.
This occurs when the maximum likelihood estimator is undefined for a given set of outcomes.
We derive the normalisation for $\tilde\delta_{\phi}$ below.

First we restate the estimator $\tilde{\delta}_\phi$ from~\autoref{eq:delta-phi-estimator-nu} for clarity
\begin{equation}
     \tilde{\delta}_\phi = \frac{1}{2}\arccos\left(\frac{1}{\eta}\frac{N_{SB}-N_{C'}}{N_{SB}+N_{C'}}\right).
\end{equation}
After $N$ measurements, the probability of obtaining $N_{SB}$ and $N_{C'}$ is
\begin{equation}
    P_\eta(N_{SB},N_{C'}|N)=\frac{N!}{N_{SB}!N_{C'}!(N-N_{SB}-N_{C'})!}P_{SB}^{N_{SB}}P_{C'}^{N_C'}(1-P_{SB}-P_{C'})^{N-N_{SB}-N_{C'}}. \label{P-nsb-nc}
\end{equation}
Using \autoref{eq:delta-phi-estimator-nu}, it is possible to write $N_{C'}$ as a function of $\tilde{\delta}_\phi$, i.e.,
\begin{equation}
    N_{C'}=N_{SB}\frac{1-\eta\cos 2\tilde{\delta}_\phi}{1+\eta\cos 2\tilde{\delta}_\phi}
\end{equation}
so that the probability of having $N_{SB}$ and $\tilde{\delta}_\phi$ after $N$ sampling measurements is
\begin{equation}
     P_\eta(N_{SB},\tilde{\delta}_\phi|N)=\frac{N!}{N_{SB}!N_{SB}\frac{1-\eta\cos 2\tilde{\delta}_\phi}{1+\eta\cos 2\tilde{\delta}_\phi}!(N-\frac{2N_{SB}}{1+\eta\cos 2\tilde{\delta}_\phi})!}P_{SB}^{N_{SB}}P_{C'}^{N_{SB}\frac{1-\eta\cos 2\tilde{\delta}_\phi}{1+\eta\cos 2\tilde{\delta}_\phi}}(1-P_{SB}-P_{C'})^{N-\frac{2N_{SB}}{1+\eta\cos 2\tilde{\delta}_\phi}}.
\end{equation}
Then, the probability of having $\tilde{\delta}_\phi$ can be obtained by considering the sum over the possible $N_{SB}$, giving
\begin{equation}
     P_\eta(\tilde{\delta}_\phi|N)= \sum_{N_{SB}}P_\eta(N_{SB},\tilde{\delta}_\phi|N).\label{P-delta-phi}
\end{equation}
This probability is not normalized. 
From~\autoref{eq:delta-phi-estimator-nu} we can see that the estimation succeeds under two conditions. The first is
\begin{equation}
    N_{SB}+N_{C'}\neq0,\label{cond1}
\end{equation}
which holds when either $N_{SB}\neq0$ or $N_{C'}\neq0$. Physically, the condition $N_{SB}=N_{C'}=0$ occurs only when the two polarization states equal to either $\ket{H}$ or $\ket{V}$. Under this condition, the relative phase loses geometrical meaning. 
The second condition is
\begin{equation}
    N_{C'}\frac{1-\eta}{1+\eta}\leq N_{SB}\leq N_{C'}\frac{1+\eta}{1-\eta},\label{cond2}
\end{equation}
linked to the fact that the higher the visibility, the more robust the estimation is against statistical fluctuations. Therefore, the probability of having a successful estimation of the relative phase is obtained by evaluating the sum of the probabilities in~\autoref{P-nsb-nc} over the values of $N_{SB}$ and $N_{C'}$ satisfying~\autoref{cond1} and\autoref{cond2}. We obtain
\begin{equation}
    P_{\mathrm{success},N}=\sum_{\substack{\mathrm{Eq.}[\ref{cond1}] \\ \mathrm{Eq.}[\ref{cond2}]}} P_\eta(N_{SB},N_{C'}|N).
\end{equation}
From the success probability, we are now able to normalize~\autoref{P-delta-phi}, by
\begin{equation}
   P_\eta(\tilde{\delta}_\phi|N)\rightarrow\frac{P_\eta(\tilde{\delta}_\phi|N)}{P_{\mathrm{success},N}} .
\end{equation}

\section{Quantum State Tomography}\label{sec:state-tomography}

\subsection{Methodology}

To perform quantum state tomography (QST) on each of the input states, we blocked one input port to the interferometer and projected the output modes of the interferometer onto each of the $X, Y, Z$ bases in turn, using the waveplates shown in \autoref{fig:experimental-setup}.
We recorded $\sim 2 \times 10^5$ detector outcomes in each basis, which were fed into a numerical maximum likelihood estimation (MLE).
The numerical MLE is framed as a constrained optimisation over positive semi-definite matrices to obtain the density matrix that maximises the likelihood of the observed outcomes.
The resulting density matrix provides the full characterisation of the input state, and the polar and azimuthal angles $\theta_i, \phi_i$ for the polarisation state of each photon were found from the corresponding mixed-state Bloch vector.
We then obtained estimates for the parameters of interest as $\tilde\theta : =\frac{1}{2}(\theta_1 + \theta_2)$ and $\tilde\delta_
\phi := \frac{1}{2}\left|\phi_1 - \phi_2\right|$.

To estimate the uncertainty for the estimates shown in \autoref{fig:asymptotic-results}, we performed a Monte-Carlo simulation assuming Poissonian noise on the outcomes in each basis.
After taking 400 Monte-Carlo samples, we found an average variance of $\text{Var}(\theta_i) = 0.002~\text{rad}$ over all measured states and $\text{Var}(\phi_i) = 0.004~\text{rad}$ over all states with $\theta \notin \{0, \pi\}$ (since $\phi_i$ is not well defined at the poles of the Bloch sphere).
The accuracy of the tomography is also limited by the extinction ratio of the PBS used to perform the measurement, which in our case is around $300:1$.
This gives an additional error on the order of $\sim1\%$, or approximately $\pm 0.01~\text{rad}$.
These errors are combined to give the error bars in \autoref{fig:asymptotic-results}.

\subsection{Precision and Bias}

\begin{figure}
    \centering
    \includegraphics[width=0.75\linewidth,
    trim= 10 16 10 0
    ]{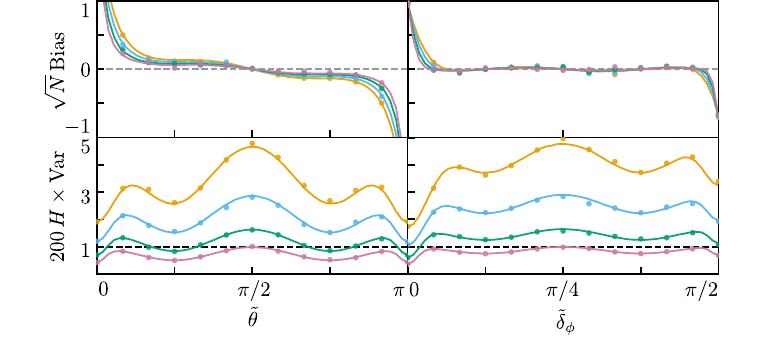}
    \caption{\textbf{Few-sample bias and precision of state tomography.} In the top row, the bias of the estimators is shown normalised by $\sqrt{N}$. 
    In the bottom row, the variance of the estimators is shown, normalised to the QCRB for $N=200$ pairs which is shown as a dashed line.
    Four values for the number of photon pairs $N \in \{120, 201, 360, 600\}$ are shown in each plot, in orange, blue, green and magenta respectively, from most biased to least, and greatest variance to least. 
    The number of pairs $N$ is always a multiple of three, since the tomography requires measurements in each of three bases.
    For the left hand plots, we choose $\phi_i = 0$ for both input photons, and sweep $\theta$.
    For the right hand plots, we choose $\theta = \pi / 2$, $\phi_1 = 0$ and sweep $\delta_\phi = \phi_2 / 2$.
    In each plot, points were calculated using MLE and lines using LS, described in the main text.
    }
    \label{fig:state-tomograph-bias-and-variance}
\end{figure}

To estimate the precision that could be achieved using state tomography, we performed another Monte-Carlo simulation at low N.
For a given input state, the outcomes in each basis will follow a binomial distribution with probabilities given by the Born rule.
We sampled this distribution, performed maximum likelihood estimation for each sample of the outcomes, and estimated the parameters as above.
We obtained variances for each parameter using 3000 Monte-Carlo samples, which are shown in \autoref{fig:state-tomograph-bias-and-variance}.

The constrained optimisation we used to find the maximum likelihood density matrix is computationally expensive.
Therefore, we also approximated MLE using the simpler least squares (LS) technique~\cite{acharya_comparative_2019}.
The matrix $M$ obtained in this way might not be physical, so we Hermitian-symmetrise the matrix $M \to \frac{1}{2}(M + M^{\dagger})$, and to ensure positive semi-definiteness, we clip eigenvalues to $\ge 0$ and renormalise to unit trace to obtain a density matrix.
We obtained variances for each parameter using 500,000 Monte-Carlo samples using LS, which are also shown in~\autoref{fig:state-tomograph-bias-and-variance}.
The variance and bias of the estimators using LS and MLE show good agreement.
For a more thorough analysis of tomographic reconstruction techniques, see Ref.~\cite{acharya_comparative_2019}.

From \autoref{fig:state-tomograph-bias-and-variance}, we see that between $\sim400$ and $\sim600$ photon pairs are required to achieve the precision given by the QCRB for $N = 200$ when performing state tomography.
This represents a two- to three-fold increase in resource usage compared to the maximum precision.
This corresponds to the resource cost of performing multiple sequential basis measurements in QST.
In the worst case scenario, two of the three basis measurements provide no information about the underlying parameters.
This appears to be the case when $\theta = \pi/2$ and $\delta_\phi = \pi/4$, for example, and three times as many photon pairs are required for the same precision.
We note that the comparison provided above represents the ideal case for QST.
In practice, additional sources of noise will also increase the variance of the QST estimators.

 \end{document}